\documentclass{an}
\usepackage{graphicx}
\usepackage{times}
\newcommand{\e}[1]{\ensuremath{{}_{\rm{#1}}}}
\newcommand{\U}[1]{\ensuremath{\mathrm{~#1}}}
\newcommand{\Myr}{\U{Myr}}
\newcommand{\kpc}{\U{kpc}}
\newcommand{\solarm}{\U{M}_{\odot}}

\begin{document}

\Pagespan{789}{}
\Yearpublication{2006}
\Yearsubmission{2005}
\Month{11}
\Volume{999}
\Issue{88}
\DOI{This.is/not.aDOI}

\title{Starburst triggered by compressive tides in galaxy mergers}

\author{Florent Renaud\inst{1}\fnmsep\inst{2}\fnmsep\thanks{Corresponding author:{renaud@astro.univie.ac.at}}
  \and
   Christian Theis\inst{1}
  \and
   Christian M. Boily\inst{2}
}
  
\titlerunning{Starburst and compressive tides}
\authorrunning{F. Renaud, Ch. Theis \& C.M. Boily}
\institute{Institut f\"ur Astronomie, Univ. of Vienna, 
T\"urkenschanzstrasse 17,
1180 Vienna, Austria
\and 
Observatoire astronomique de Strasbourg and CNRS UMR 7550,
11 rue de l'Universit\'e
67000 Strasbourg, France
}

\received{20 August 2008}
%\accepted{11 Nov 2005}
%\publonline{later}


\abstract{The tidal field of galaxies is known generally to be disruptive. However, in the case of galaxy mergers, a 
 compressive mode of tidal wave may develop and last long enough to cocoon the formation of star clusters. Using an N-body simulation of 
the Antennae galaxies, we derive the positions of these compressive regions and the statistics of their duration. 
Excellent agreement between the spatial distribution of tides and observed young clusters is found, while the characteristic e-folding 
times of $10$ to $30$ Myrs derived for the tidal field compare well with cluster formation time-scales.}

\maketitle

%%%%%%%%%%%%%%%%%%%%%%%%%%%%%%%%%%%%%%%
\section{Introduction}
It is well known that gravitational tides work  to disturb or even destroy galactic disks and substructures. However, one should not forget 
that tidal fields can also be {\it compressive} and have the opposite effect. For example, the tidal 
force perpendicular to the line connecting two spherical mass distributions is compressive.
 Moreover, for specific mass distributions, the tide can be compressive in {\it all} directions (e.g. 
Valluri~1993). When that is the case, a tidal force will work to gather matter and keep it bound. 
A picture emerges from such considerations whereby  e.g. a molecular cloud experiencing 
 a compressive tide for a  long-enough time would collapse, form a cluster of stars, and keep the gas in the vicinity of the
 proto-cluster: stars would form even in  a globally low star formation efficiency environment. Furthermore, their feedback (e.g.,
 stellar winds or radiation) would not affect the cluster as much as if it were in isolation (Elme-green \& Efremov~1997), owing to additional binding energy. 
It is possible  that this effect would significantly alter the mass-loss 
process which likely affects the late stage of the formation of star clusters (e.g. Bastian \& Goodwin~20-06). 
A  cluster which forms within a compressive tidal field would be super-virial after it relaxed to equilibrium, and thus would 
 re-expand or even dissolve whenever the tide switches from a compressive,  to an extensive mode. 
This could be a clue to understand the process driving the rapid dissolution (or, ``infant-mortality'')  often invoked 
in relation to the demographics of young clusters in the Antennae galaxies (Whitmore et al.~2007).

Our goal here is to show that compressive tidal modes are effectively operative on a cluster-formation time-scale. 
It is therefore likely that clusters are super-virial in their infancy but are prevented from dissolving 
by the background tide. 
After a brief presentation of compressive tides, we pre-sent an N-body model of the Antennae galaxies and the 
statistics of their tidal field.

%%%%%%%%%%%%%%%%%%%%%%%%%%%%%%%%%%%%%%%
\section{Tidal tensor and compressive tides}
Tides are usually evoked for their destructive effects. In the case of the  merger of two spiral 
galaxies, tidal bridges and tails commonly form from the extension of the 
disks material. This disruption can be described by the tidal tensor
\begin{equation}
\label{tidal_tensor}
T^{ij} \equiv -\partial_i \left(\partial_j \phi \right)
\end{equation}
where $\phi$ is the gravitational potential. Because of its symmetry, it can be set in an orthogonal form. Its eigenvalues then represent either a compression (if negative) or an extension (if positive) along the corresponding eigenvector. 
Consequently, inspection of the sign of the maximum eigenvalue allows to distinguish between fully compressive and extensive modes.

\begin{figure} 
\includegraphics[width=\linewidth]{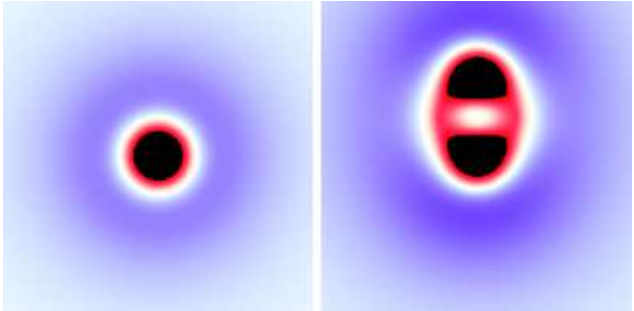}
\caption{Totally compressive (red) and extensive (blue) tidal regions for the  logarithmic potential in isolation (left-hand panel), 
and for two 
identical logarithmic  potentials, but one shifted 
by $\sim 2b$ units
along the vertical axis (right-hand panel).}
\label{fig:log_pot} 
\end{figure}

For the  logarithmic potential 
\begin{equation}
\label{log_pot}
\phi\e{log}(\mathbf{x}) = v_c^2 \ln{\sqrt{1+\frac{x_kx^k}{b^2}}}
\end{equation}
(we apply Einstein's summation convention), the tidal tensor reads
\begin{equation}
\label{log_t}
T^{ij}\e{log}(\mathbf{x}) = -v_c^2 \frac{\delta^{ij}\left( b^2 + x_kx^k \right) -2 x_ix_j}{\left( b^2 + x_kx^k \right)^2}
\end{equation}
and produces compressive tides when $\sqrt{x_kx^k} = r<b$, i.e. in the inner part of the potential. Adding a second, similar potential (same 
parameters, as for a  major galaxy merger) shifted along the vertical axis, we show on Fig.~\ref{fig:log_pot} (right panel) how the 
compressive regions 
(red on the figure) covers a larger volume than for a potential taken in isolation (left panel on that figure).

%%%%%%%%%%%%%%%%%%%%%%%%%%%%%%%%%%%%%%%
\section{The Antennae}
\begin{figure}
\includegraphics[width=\linewidth]{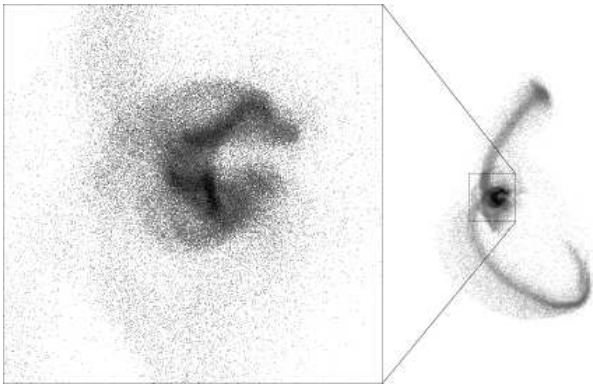} 
\caption{Column density of the N-body model of the Antennae galaxies and a zoom-in on the central region ($\sim 40 \kpc$ wide) at the present time ($300 \Myr$ after the first pericenter passage).} 
\label{fig:antennae} 
\end{figure}

We applied the tensor analysis 
to an N-body simulation of the Antennae galaxies (NGC4038/39). As we consider the gravitational field only, no gas 
was included in the simulation. This model uses an  S0 and an Sa progenitors, each made of an exponential disk, a Hernquist bulge and an 
iso-thermal dark-matter halo (see Renaud et al. 2008 for more details). The progenitor galaxies have a mass ratio of 1:1 
for a total mass (stars + 
dark matter) of $\sim 9.2 \times 10^{10} \solarm$. The galaxies were set on a prograde elliptical orbit ($e \approx 0.96$) with an initial 
separation of $75 \kpc$. We obtained a good match to the observations, both in terms of morphology and the velocity field, 
some $300 \Myr$ after 
the first pericenter passage (Fig.~\ref{fig:antennae}). The mass resolution of $\sim 2 \times 10^5 \solarm$ means that a single 
particle is a cluster-size element.

\begin{figure}
\includegraphics[width=\linewidth]{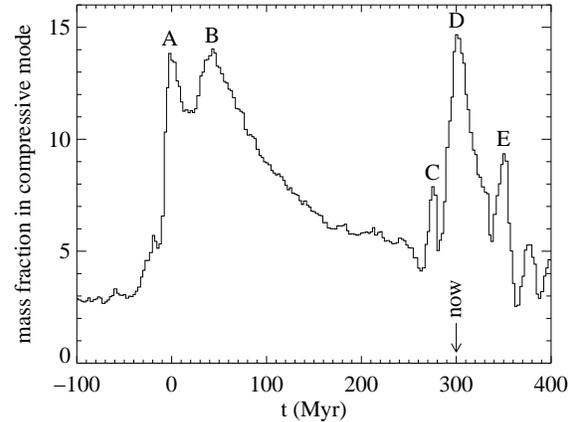}
\caption{Percentage of the mass of the disks in compressive mode as a function of time. Peaks A and B coincide with the first pericenter 
passage while C and D are to be linked with the second one. Other small spikes would be rapid evolution of tidal clumps falling onto the 
core.} 
\label{fig:histo}
\end{figure}

We extracted the tidal tensors of all 400,000 disks particles for a large sequence of simulation snapshots to achieve 
a time-interval as low as $2.5 \Myr$. 
 Fig.~\ref{fig:histo} plots the fractional number of particles (in percentage) in a compressive tidal field as a 
function of time. 
During the early stages of the merger, the mass fraction stays 
at a level of 3\% (as for each progenitor taken in isolation) until  
it suddenly increases at the first pericenter passage (peaks A and B). 
The fraction rapidly drops and almost  returns to its initial value  as the system re-expands and 
the progenitors move apart. 
However,  it rises again by a factor $\simeq 5$ at the second passage (peaks C and D, the current time, labeled 'now'). 
Note that this increase 
compares well with typical star formation rates  
obtained from hydrodynamical simulations of starburst mo-dels with  similar setups (see Di~Matteo et al. 2008). 
We conclude from the simulation run that stars and star clusters 
formed $\sim 300 \Myr$ ago are to be linked with the first pericenter passage and the creation of the two tidal tails.

\begin{figure}
\includegraphics[width=\linewidth]{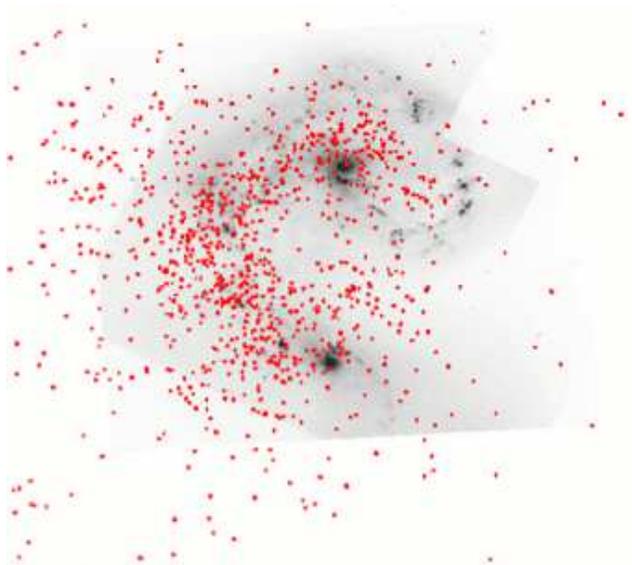} 
\caption{Spatial distribution of the compressive regions in the central part (red dots), 
laid over an optical image  of the Antennae (grey scale background, adapted from Mengel et al.~2005). 
The data points displayed correspond to the current time (peak D on Fig.~\ref{fig:histo}).} 
\label{fig:map}
\end{figure}

Fig.~\ref{fig:map} maps the position of these compressive regions at the present time (peak D on Fig.~\ref{fig:histo}) to 
the Antennea galaxies. The background is 
a composite HST/VLT image adapted from Mengel et al.~(2005), showing the nuclei and the overlap region. 
Note the good match between the 
particles in compressive mode and the sites of cluster formation, mainly on the bow shape around the northern nucleus (where north is up) 
and in the overlap 
region, between the two cores. This is a good hint that the tidal field plays an important role in the formation of such stellar structures.

\begin{figure}
\includegraphics[width=\linewidth]{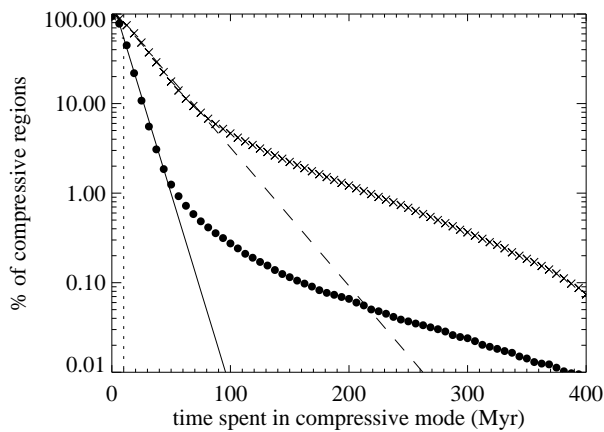} 
\caption{Distribution of time-intervals spent in compressive mode by the disk particles. The circles show the distribution of 
uninterrupted sequences. This is well-fitted by an exponential law (solid curve) with characteristic time-scale of 10 Myr up to the first 
100 Myr. The crosses and dashed line show the total time accrued in compressive mode (with possible on-off episodes). 
The associated time-scale is 30 Myr.} 
\label{fig:period}
\end{figure}

To backup  this idea, we plot on Fig.~\ref{fig:period} the mass fraction in compressive modes as a function of their duration. 
We show that about 55\% of the disk mass experience a continuous tidal compression lasting longer than 10 Myr. These figures 
overlap with age estimates of young clusters of the Antennae, and thus emphasize the role of the compressive tidal mode in the 
formation of these systems.

%%%%%%%%%%%%%%%%%%%%%%%%%%%%%%%%%%%%%%%
\section{Conclusions}
Using a gravitational N-body simulation of the Antennae galaxies, we derived the statistics of cluster-size elements experiencing a tidal 
compression. The results show a good match with observational data, both in terms of spatial distribution and characteristic times. 
The statistics of tidal field evolves rapidly 
as the number of compressive regions increases by a factor 5 during the merger. The characteristic duration of compressive
 modes and their spatial distribution suggest that they play an important role for 
 the formation of clusters, as they would  keep them  bound and prevent
 or delay their rapid dissolution. A follow-up study will explore the response of gas on a cluster scale when considering the external 
tidal field as boundary conditions.

\acknowledgements
FR acknowledges a scholarship from the IK\-~I033-N \emph{Cosmic Matter Circuit} at the University of Vienna. CT is grateful to the German Science Foundation (DFG) for financial support within the priority program 1177.

%%%%%%%%%%%%%%%%%%%%%%%%%%%%%%%%%%%%%%%%%%%%%%%%%%%%%%


\begin{thebibliography}{}
\bibitem[Bastian \& Goodwin(2006)]{Bastian06} Bastian, N., \& Goodwin, S.~P.\ 2006, MNRAS, 369, L9
\bibitem[Di Matteo et al.(2008)]{DiMatteo08} Di Matteo, P., Bournaud, F., Martig, M., Combes, F., Melchior, A.-L., \& Semelin, B.\ 2008, AAP, submitted
\bibitem[Elmegreen \& Efremov(1997)]{Elmegreen97} Elmegreen, B.~G., \& Efremov, Y.~N.\ 1997, ApJ, 480, 235
\bibitem[Mengel et al.(2005)]{Mengel05} Mengel, S., Lehnert, M.~D., Thatte, N., \& Genzel, R.\ 2005, AAP, 443, 41
\bibitem[]{Renaud08} Renaud, F., Boily, C.M., Fleck, J.-J., Naab, T., Theis, C.: 2008, MNRAS Letters, accepted 15 September 
\bibitem[Valluri(1993)]{Valluri93} Valluri, M.\ 1993, ApJ, 408, 57
\bibitem[Whitmore et al.(2007)]{Whitmore07} Whitmore, B.~C., Chandar, R., \& Fall, S.~M.\ 2007, AJ, 133, 1067
\end{thebibliography}
\end{document}